\documentclass[aps,nofootinbib]{revtex4}
\def\be{\begin{equation}}
\def\ee{\end{equation}}
\def\ba{\begin{eqnarray}}
\def\ea{\end{eqnarray}}

\usepackage{amsmath,bbm}
\usepackage{amssymb}
\usepackage{mathrsfs}
\usepackage{mathtools}
\usepackage{amsthm}
\usepackage{xcolor}
\usepackage{verbatim}
\usepackage[pdftex]{graphicx}
\usepackage{hyperref}
\hypersetup{
	colorlinks=true,
	linkcolor=blue,
	filecolor=magenta,      
	urlcolor=cyan,
}

\DeclareFontFamily{U}{rsfs}{}         
\DeclareFontShape{U}{rsfs}{m}{n}{<5> rsfs5 <6><7> rsfs7          %
  <8><9><10><10.95><12><14.4><17.28><20.74><24.88> rsfs10}{}     %
\DeclareMathAlphabet{\mathfs}{U}{rsfs}{m}{n}                     %
 \usepackage{braket}

\newcommand{\RN}[1]{%
  \textup{\uppercase\expandafter{\romannumeral#1}}%
}

\newcommand{\va}{\scriptscriptstyle}

\usepackage{braket}
\newcommand{\sbraket}[1]{\braket{\!\braket{#1} \!}}

\begin{document}

\title{Is Planckian discreteness observable in cosmology?}

\author{Gabriel R. Bengochea,}
\email{gabriel@iafe.uba.ar}
\affiliation{Instituto de Astronom\'{\i}a y F\'{\i}sica del Espacio (IAFE),
CONICET, Universidad de Buenos Aires, (1428) Buenos Aires, Argentina.}

\author{Gabriel Le\'{o}n}
\email{gleon@fcaglp.unlp.edu.ar}
\affiliation{Grupo de Cosmolog\'{\i}a, Facultad de Ciencias Astron\'omicas y Geof\'{\i}sicas,
Universidad Nacional de La Plata, Paseo del Bosque S/N 1900 La Plata, Argentina.
CONICET, Godoy Cruz 2290, 1425 Ciudad Aut\'onoma de Buenos Aires, Argentina.}

\author{Alejandro Perez}
\email{perez@cpt.univ-mrs.fr}
\affiliation{Aix Marseille Universit\'e, Universit\'e de Toulon, CNRS, Centre de Physique Th\'eorique, 13000 Marseille, France.}

\vskip3cm

\begin{abstract}
A Planck scale inflationary era---in a quantum gravity theory predicting discreteness of quantum geometry at the fundamental scale---produces the scale invariant spectrum of inhomogeneities with very small tensor-to-scalar ratio of perturbations 
and a hot big bang leading to a natural dark matter genesis
scenario. Here we evoke the possibility that some of the major puzzles in cosmology would have an explanation rooted in quantum gravity.


\end{abstract}
 
\maketitle

\section{Introduction}

Standard inflationary models of cosmology are successful in reproducing some of the qualitative features observed in the cosmic microwave background (CMB) \cite{Planck:2018jri}. This success relies on three key assumptions which, we will argue here, can be changed without altering the salient features of the observable outcome. We propose a paradigm shift where quantum gravity is at the heart of the mechanism leading to structure formation in our universe, and observe that the new perspective could alleviate existent tensions and produce Planckian physics verifiable predictions. 

The first assumption of standard approaches is the introduction of an intermediate scale---the inflationary scale, which, in order of magnitude, can be identified with the reheating temperature---between those scales appearing in the standard model of particle physics and the Planck scale. Instead, we will see that it is consistent to assume that only the Planck scale plays a role in pre big bang cosmology bypassing constraints which the standard inflationary scenario derives from the lack of observation of gravitational waves in the CMB \cite{BICEP2:2018kqh}. The second assumption is that of the validity of the standard quantum field theory at energy scales that are much higher than the Planck scale (the so-called trans-Planckian problem). Instead, here we will assume that the fundamental dynamics is described by a fundamentally discrete quantum theory of gravity imposing a physical cutoff at the fundamental scale. Finally, there is a unitarity-violating\footnote{Here we refer to the fact that information is lost in the process where the universe goes from a quantum description to a classical one whose observable features do not allow for the reconstruction of the initial state. In the standard inflationary paradigm, the process by which the universe becomes what we observe at present times is associated to either some form of decoherence \cite{Kiefer:1998qe, Polarski:1995jg}, or some effective measurement type of transition \cite{Perez:2005gh, Martin:2012pea}. In all cases the past state of the universe cannot be reconstructed from the knowledge of its final observable features only; in violation of the expectations from a unitary fundamental description. } symmetry breaking involved in the standard inflationary paradigm when quantum fluctuations in a vacuum-like state of the inflaton---perfectly homogeneous and isotropic as a quantum state---become the classical inhomogeneities of the energy density.  Such symmetry breaking from a homogeneous quantum state to an inhomogeneous classical configurations is often referred to as the quantum-to-classical transition.  Instead, here we propose that inhomogeneities are present from the beginning in the fundamental structure of quantum geometry at the Planck scale\footnote{A generic prediction of most grand unified theories (GUTs) is the  production of stable magnetic monopoles at the GUT scale, $E_{\rm GUT}$ \cite{tHooft:1974kcl}. The standard inflationary paradigm resolves this issue by postulating that inflation occurs at an energy scale $E_{\rm inf} \ll E_{\rm GUT}$. Consequently, the exponential expansion dilutes any pre‑existing magnetic monopoles; furthermore, if the reheating scale also satisfies $E_{\rm reh} \ll E_{\rm GUT}$, no new monopoles are produced after inflation. If this were the case, our model (which occurs at the Planck scale) would not allow the dilution of such monopoles. This conclusion depends, of course, on the validity of the GUT framework, which—though highly elegant—relies on several assumptions. Notably, current bounds on proton decay appear to favor supersymmetric extensions or more refined model constructions that rely on additional assumptions (all of which remain hypothetical at present). Here, we explore a perspective in which monopoles may not be produced at energy scales below the Planck scale in the eventual extension of the standard model of particle physics.} and simply percolate as semiclassical inhomogeneities in the matter fields during cosmological dynamics without any violation of unitarity (there is no quantum to classical transition and the evolution of the universe can in principle be entirely described quantum mechanically and approximated semiclassically). 

The paper is organized as follows: We start with section \ref{sect1}, where we develop the ideas of the model and its connection to the emergence of inhomogeneities from a fundamental Planckian granularity. In section \ref{sect2} we show the expected scalar and tensor spectra for primordial perturbations in the proposed framework and in section \ref{sect3} we discuss the possibility that primordial black holes of Planckian mass constitute dark matter. Finally, in section \ref{conclusions}, we present our conclusions. Throughout this work, we will use a $(-,+,+,+)$ signature for the spacetime metric and units where $c=1=\hbar$.

\section{Inhomogeneities from Planckian Discreteness}\label{sect1}

Inhomogeneities observed in the CMB are, in the view we propose, simply the stretched out traces of the built in fundamental granularity of spacetime geometry at the Planck scale.

To make the idea precise, we describe the background geometry of the universe at large scales by a flat Friedmann-Lema\^{\i}tre-Robertson-Walker (FLRW) metric
\be\label{bground}
ds^2=-dt^2+a(t)^2 d\vec x\cdot d\vec x,
\ee
with $a(t)$ the scale factor.
We assume that before reheating the universe is in an inflationary 
phase driven by an approximately constant Planck scale energy density, acting temporarily like a cosmological constant. This implies $a(t)=a_0 \exp(H t)$, with the Hubble rate $H\approx m_p$. Such primordial dark energy is expected to decay and produce the reheating of the universe that leads to a hot (post inflationary) big-bang. We are not concerned here with the details of the reheating phase; however, it is easy to see that, in the present context, a rapid reheating leads to a hot big bang with an initial temperature close to the Planck temperature.

In the standard inflationary paradigm the state of perturbations of the inflaton field is assumed to be in a vacuum state (say the Bunch-Davies vacuum), 
$\ket{0}$, where $\braket{0|\widehat {\delta \phi_k}|0}=0$. That state does not break the FLRW symmetries in the sense that the expectation value
of any local operator is homogeneous and isotropic. In particular, the expectation value of the corresponding energy-momentum tensor is homogeneous and isotropic, i.e., 
\be\label{two} \braket{0|\widehat T_{ab}(\vec x, t)|0}=\braket{0|\widehat T_{ab}(\vec x+\vec r, t)|0}\ee for arbitrary $\vec r$. The actual state of the universe at later times 
is not homogeneous nor isotropic which demands a symmetry breaking---referred to as the quantum-to-classical transition. There is no
consensus about how such a transition actually happens in the standard picture \cite{Hartle:2019hae, Berjon:2020vdv}.

In contrast, here we propose that the symmetry is broken already in the primordial phase at the microscopic scale due to the fundamental Planckian granularity of quantum gravity. The constraints from the requirement of low energy Lorentz invariance \cite{Perez:2003un, Collins:2004bp, Collins:2006bw} suggest\footnote{How can the discreteness scale appear in the context of quantum gravity without conflicting the relativity principle at low energies? Observables in a generally covariant gravitational context are relational which imply that, in order to measure or interact with the fundamental scale, the associated degrees of freedom must work as the Einstein's rods and clocks in terms of which the geometry is operationally defined. To serve as measuring notions, such degrees of freedom must break scale invariance. The simplest case is that of a field with mass $m$ where excitations define a particular proper rest frame where, not only the energy scale $m$ has an invariant meaning as the rest mass of the excitation, but also the Planck scale acquires its invariant  meaning in relation to that dynamical frame. In turn photons, an emblematic low energy scale-invariant degree of freedom, cannot measure (interact with) discreteness as they cannot select any particular rest frame with respect to which a fundamental scale would be meaningful: it simply does not work as a good rod and clock relational excitation to interact with the microscopic dynamical granularity.} that only 
matter fields that break scale invariance interact with the Planckian granularity \cite{Perez:2018wlo, Perez:2017krv}. The matter content in the standard model of particle physics suggests 
that at high energies only scalar fields like the Higgs can break scale invariance. 
Thus, the quantum state describing the scalar field perturbations cannot be close to anything like a vacuum state near the fundamental scale, 
 $\ket{\psi}\not=\ket{0}$, but  rather given by a state with non trivial inhomogeneities induced by the grainy structure of quantum geometry. This is the key departure of our proposal from the standard inflationary paradigm. 
 
A precise description  of the fundamental mechanism for the generation of the inhomogeneities would require understanding of the quantum gravity dynamics at the Planck scale. Here, we take a phenomenological approach based on the assumption (confirmed by the study of certain simplified models \cite{Ashtekar:2011ni}) that a mean-field semiclassical approximation can become a good approximation close to the fundamental scale. The perspective we adopt is similar to the one taken in the Ginsburg-Landau phenomenological theory of superconductivity, where the dynamics of an emergent collective variable is modelled according to very general principles. The emergent collective variable of interest here is the expectation value of the scalar field perturbations (which we assume is the single scale-invariance breaking degree of freedom; the Higgs or another scalar). These perturbations are created from the interaction with a quantum geometry whose grainy structure breaks the symmetries of the FLRW background. 
Concretely, the mean-field-variables of interest are given by the Fourier modes of the scalar perturbations \be\label{semi}
\delta \phi_k\equiv \braket{\psi|\widehat {\delta \phi_k}|\psi}\not=0, \ \ \ k\le m_p a, 
\ee
which, in the absence of interactions, follow the standard semiclassical evolution equations 
\be\label{km}
 \delta \ddot\phi_k+3 H \delta\dot \phi_k+\frac{k^2}{a^2} \delta \phi_k
 =0,
\ee
where $H$ is the (approximately) constant Hubble rate
\footnote{\label{footy}For simplicity we neglect here possible self interaction terms by taking ${d^2V(\phi_0)}/{d \phi^2}\ll H^2$,     
{in addition,  metric perturbations that generally appear in this equation are absent during the De Sitter phase as a consequence of the Weinberg theorem \cite{Weinberg:2003sw, Weinberg:2008zzc}. We will do the same in equation \eqref{hijeom}.}}.
For modes with $k^2< a^2 H^2$ the friction term dominates and the 
solutions freeze out exponentially fast at horizon crossing. 

As mentioned, we assume that the horizon scale $H\approx m_p$: the primordial vacuum energy starts at the natural Planck value. In an approach to quantum gravity where physics is discrete at the fundamental scale no spacetime symmetries can be preserved microscopically: the state of scalar perturbations will emerge asymmetric from the Planck scale.
Perturbations are born at the Planck scale $k=a m_p$ and, once stretched by the de Sitter expansion, decouple from the discrete microscopic structure that generated them to propagate freely according to the mean field evolution equations \eqref{km} which freezes the inhomogeneities produced once the modes expand to wavelengths longer than the Planck scale.

In the absence of a precise microscopic theoretical understanding, we resort to statistical methods akin to Einstein's stochastic depiction of Brownian motion.
Thus, we assume that the scalar field is excited in a random fashion and we model the situation by a stochastic process whose ensemble averages are denoted
with the double-brakets $\sbraket{}$ (not to be confused with quantum mechanical state-averages). The first moments of the stochastic process is taken as vanishing, namely 
\be\label{stochyk}
 \sbraket{{\delta \phi_k}}=0.
\ee
We assume that the stochastic process respects the symmetries of the FLRW background, which implies that the second moments 
must be given by  
\be\label{stochyk}
 \sbraket{{\delta \phi_k}{\delta \phi_q} }= P_{\delta\phi} (k)\  \delta(\vec k+\vec q),
\ee
where $P_{\delta\phi} (k)$ is the so-called power spectrum of the scalar perturbations.
Fourier transforming the previous equation it follows that \cite{Peter:1208401, Brandenberger:2003vk} 
\be\label{keyly}
\sbraket{\delta\phi(t,\vec x)\delta\phi(t,\vec x)}=\frac{1}{(2\pi)^3} \int \limits_\mu^{a(t) m_p} dk^3 P_{\delta \phi} (k),
\ee
where the lower limit of integration $\mu$ is an infrared cutoff---the size of the universe patch where the background metric of Eq. \eqref{bground} is a good approximation---while the upper limit at $k_{max}=a(t) m_p$ implies that perturbations are born when their physical wavelength corresponds to the Planck scale. This represents the process of creation of inhomogeneities via the interaction with the microscopic granular structure at the fundamental scale. As soon as created the modes cross the Hubble horizon and get frozen, thus the relevant time dependence on the generation of field perturbations is encoded in the time dependence of the upper integration limit of the previous equation\footnote{The assumption that the semiclassical equations \eqref{km}---as well as Einstein equations---are valid appears as a strong hypothesis at first sight. However, notice that as soon as created at the Planck scale, the perturbations are quickly exponentially stretched into the long wavelength regime where semiclassical equations hold (long with respect with the Planck scale in the co-moving frame). In this sense, we are simply encoding the deviations from semiclassicality (whose precise features can only be described in a full quantum treatment) in the stochasticity of the generation process we postulate.}. This time dependence requires energy to flow from the Planckian microscopic substratum into the scalar field perturbations: as the Planckian granularity is continuously exciting the quantum state of the scalar field with inhomogeneities \eqref{semi}, the energy (injected into the scalar field perturbations) is encoded in a non trivial energy-momentum tensor violating FLRW symmetries (i.e. Eq. \eqref{two} is violated), which we model as
\begin{equation}\label{Tabnueva}
	\nabla^a \sbraket{T_{ab}} = \sbraket{\xi \nabla_b \phi} 
\end{equation}
where $\xi$ is a stochastic scalar variable with vanishing first moment.

We conclude this section by mentioning an alternative mechanism for generating primordial inhomogeneities that, like our model, does not rely on vacuum fluctuations. In Ref. \cite{Brahma:2021tkh}, a model based in matrix theory \cite{Banks:1996vh} and motivated by string‑gas cosmology is explored. There, the early Universe is assumed to begin in a thermal (rather than vacuum) state, so that thermal fluctuations seed the primordial perturbations. These thermal fluctuations yield an approximately scale‑invariant spectrum of cosmological perturbations and gravitational waves. Although the tensor‑to‑scalar ratio is expected to be within the observable range, its exact value has yet to be determined theoretically. Our framework likewise predicts a nearly scale‑invariant spectra; however, as we show in the following section, we expect a strongly suppressed amplitude for primordial gravitational waves.

\section{Scalar and Tensor Power Spectra}\label{sect2}

In the last section, we saw that within the framework of the model a flow of energy from the spacetime granularity into the scalar field perturbations must occur, and that such energy is encoded in a non trivial energy-momentum tensor. One can estimate the energy cost via the ensemble average of the energy-momentum tensor which takes the perfect fluid form, \ba\label{noname}
\!\!\! \sbraket{T_{ab}[\delta\phi]}&=&\sbraket{\nabla_a \delta\phi \nabla_b\delta\phi}-\frac{g_{ab}}{2} \sbraket{\nabla_\alpha\delta\phi \nabla^\alpha\delta\phi} 
\approx \underbrace{\frac{\sbraket{(\vec\nabla\delta\phi)^2}}{2a^2}  }_{\equiv \rho^{(2)} }\  u_au_b \underbrace{-\frac{\sbraket{(\vec\nabla\delta\phi)^2}}{6a^2} }_{\equiv P^{(2)} }\ h_{ab},
\ea
due to the homogeneity and isotropy of the stochastic generation process %
\footnote{We first focus on  the quantity $\sbraket{\nabla_a \delta\phi \nabla_b\delta\phi}$, which, 
from isotropy of the stochastic process, can be written as:
\be\label{tab}
\sbraket{\nabla_a \delta\phi \nabla_b\delta\phi}=\sbraket{\delta\dot \phi^2} u_au_b+ \frac{1}{3 a^2}\sbraket{\vec \nabla \delta\phi\cdot \vec \nabla \delta\phi} h_{ab}\approx \frac{1}{3 a^2}\sbraket{\vec \nabla \delta\phi\cdot \vec \nabla \delta\phi} h_{ab},
\ee
where $h_{ab} = g_{ab}+u_au_b$  is the space metric and $u^a=\partial_t^a$ is the 4-velocity of the comoving frame and we have neglected the 
$\sbraket{\delta\dot \phi^2}\ll \sbraket{\vec \nabla \delta\phi\cdot \vec \nabla \delta\phi}$.
With the previous result $\sbraket{T_{ab}[\delta\phi]}$ becomes as in \eqref{noname}.}
.
Energy balance is understood via the work per unit time necessary to produce the inhomogeneities \be \frac{dW}{dt}=u^a\nabla_a W\equiv -\nabla^a \sbraket{T_{ab}} u^b\ee which leads to the left hand side of a continuity equation
\ba\label{kiko}
\frac{dW}{dt} [P_{\delta\phi}] &\equiv& \frac{d}{dt}\rho^{(2)} +3 H \left(\rho^{(2)}+P^{(2)}\right)
=\frac{d}{dt}\rho^{(2)} +2 H \rho^{(2)}=J,
\ea
where in the last equality we have used the equation of state for the perturbations that follow from Eq. \eqref{noname}, and we introduced the current $J$, on the right hand side, representing the injection of energy associated with the production mechanism of inhomogeneities. This is, we have defined $J \equiv \sbraket{ \xi \nabla_a \phi } u^a$ in \eqref{Tabnueva}.

We assume that mode excitations in the scalar field are created randomly when the physical wave length of the mode coincides with the Planck scale, namely when $k=a m_p$.
This assumption relates $\rho^{(2)}$ and the power spectrum in Eq. (\ref{stochyk}) according to 
\ba\label{54}
\rho^{(2)} [P_{\delta\phi}]\equiv \frac{1}{2a^2} \sbraket{(\vec\nabla\delta\phi)^2}&\approx&\frac1{2\pi^2}\int\limits_{\mu}^{a(t) m_p} dk k^2  \left(\frac{k^2}{2 a^2}\right)P_{\delta\phi}(k),
\ea
where the upper limit in the integration realizes the idea that modes are created at the Planck scale. 
The creation of the modes by the Brownian interaction with the granularity costs energy encoded in the current $J$ in Eq. \eqref{kiko}.
As the only energy scale present during the inflationary phase is $H$, dimensional analysis implies that  $J=\gamma H^5$,
where $\gamma$ is a dimensionless parameter \footnote{ It appears natural to associate $\gamma$ with an order parameter of the violation of scale invariance. For instance $\gamma\approx m/m_p$ for a massive field with mass $m$. When the present scenario is applied to the Higgs scalar  this identification leads to the correct magnitude for the power spectrum of scalar perturbations \cite{Amadei:2021aqd}.}.  
Using that $dt=da/\dot a=H da/a$ we can rewrite Eq. \eqref{stochyk} as
\ba\label{condi}
\frac{d}{da} W^{\rm pert.}[P_{\delta\phi}(k)]\equiv \frac{d \sbraket{\rho^{(2)}}}{da}+\frac{2}{a} \sbraket{\rho^{(2)}}
&=&\gamma \frac{H^4}{a}.\ea
The previous energy-balance equation---the analog of Einstein's detail balance equations in the context of Brownian motion---uniquely determine the power spectrum $P_{\delta\phi}(k)$ as follows.
First consider the term ${d \sbraket{\rho^{(2)}}}/{da}$ in the previous equation in view of Eq. \eqref{54}. There are two contributions: a term coming from the derivative with respect to $a$ on the integration limit (which gives $H$ times the integrand evaluated at $k=a H$), and the integration of the $a$-derivative of the integrand.  It turns out that the second contribution 
cancels exactly with the term $\frac{2}{a} \sbraket{\rho^{(2)}}$ in Eq. \eqref{condi}. The reason is that once created we assume that the fluctuations evolve according the semiclassical scalar field equations \eqref{km} that imply  that energy is conserved. Thus, Eq. \eqref{condi} reduces to:
\ba
\left. H k^2  \left(\frac{k^2}{2 a^2}\right)P_{\delta\phi}(k)\right|_{k=aH}= \frac{ \gamma H^4}{a}\label{kirikiri}
\ea
Whose solution is the scale invariant spectrum $$P_{\delta\phi}(k)=2\gamma \frac{H^2}{k^3}.$$ The power spectrum of scalar perturbations 
can be translated into the (gauge invariant) power spectrum curvature perturbations observed in the CMB, where small deviations of scale invariance (as the one observed \cite{Planck:2018jri}) 
are associated to details of the evolution of the Hubble rate during inflation, where a minimal departure from a perfect de Sitter type of expansion should be considered. Such more detailed analysis restricts dimensionless parameter $\gamma< 10^{-10}$ \cite{Amadei:2021aqd}.

As photons, gravitons are not suitable scale breaking degrees of freedom that could directly interact with the fundamental discreteness.
However, the production of inhomogeneities via the interaction of the scalar matter field with the granularity at horizon crossing is a time dependent process that should subsequently lead to the production of gravitational waves. 
In the semiclassical treatment the metric perturbations are sourced by the quantum expectation values of the energy-momentum tensor.  At first-order, the tensor perturbations of the metric $h^{(1)}_{ij}$---i.e. the gravitational waves (GW)---do not contain matter sources, so  $h^{(1)}_{ij} =0$. However, the semiclassical Einstein's equations at second order yield the following equation for the GW, 
\begin{eqnarray}\label{hijeom}
 \ddot h_{ij}^{(2)} -\frac{\nabla^2 h_{ij}^{(2)}}{a^2} +3  \frac{\dot a}{a} \dot h_{ij}^{(2)} = \frac {\left\{ 
32\pi G \partial_i\delta \phi \partial_j \delta\phi \right\}^{\va TT}}{a^2}. 
\end{eqnarray}
where $\{ ...  \}^{TT}$ denotes the transverse and traceless part of $\{...\}$ \footnote{As explained in footnote \ref{footy}, first order metric perturbations can be neglected during the De Sitter phase.}.
The previous equation implies that the spectrum of the primordial GW quadratic in the matter perturbations. A precise analysis shows \cite{gabrieles} that the tensor power spectrum is scale invariant with $$P_h (k) \simeq \gamma^2 \frac{H^4}{m_p^4 k^3}.$$ The quadratic dependence on $\gamma$ suggests the strong suppression we evoke above.  A precise calculation of the tensor-to-scalar ratio $r$ requires depends on details of the deviations from De Sitter. Nevertheless, detail analysis \cite{gabrieles} shows that $r<10^{-10}$.

\section{The Gravitational Miracle}\label{sect3}

After the inflationary phase, the effective vacuum energy of the order of the Planck density that was driving it decays via interactions with matter degrees of freedom reheating the universe up to  around the Planck temperature. Such a hot big-bang will produce via thermal fluctuations Planckian mass primordial black holes (PBH). There are reasons to expect that discreteness will have a dramatic effect in the physical properties of such elementary PBHs and that, contrary to the semiclassical expectation that they would evaporate via Hawking radiation (an expectation only justified for large black holes in Planck unites), black holes of Planckian mass could be stable and interact only gravitationally  \footnote{One has to keep in mind that 
at the Planck scale the very notion of smooth geometry is expected to be lost and that the particles postulated here might have properties very different from 
the macroscopic BH solutions which Hawking radiate. A Planck mass particle is most natural from the perspective of quantum gravity, it could be stable due to 
quantum dynamical bouncing \cite{Rovelli:2018okm} or by being effectively extremal via quantum gravity effects (see \cite{Hayward:2005gi, Bodendorfer:2019cyv, Baranov:2024myo} and the very general family of models analysed in \cite{Munch:2022teq}) or if with spin of the order of $\hbar$ (see for instance \cite{Carneiro:2021rwm}). This last conservative possibility is attractive as such particles would be Fermions.}. If so their abundance can be easily estimated and the result is striking as it corresponds, in order of magnitude, to the amount needed to explain the dark-matter density today. As we will see in more detail below, the only hypothesis that goes into this estimate is that Planck mass stable particles are part of the spectrum of quantum gravity and that they interact only gravitationally. In analogy to a similar coincidence, called the WIMP miracle
in the weakly interacting sector of supersymmetric extensions of the standard model,  we term this coincidence the {\em gravitational miracle} \cite{Amadei:2021aqd, Barrau:2019cuo}.  As these particles only interact gravitationally, the form of dark matter that follows from these assumptions---requiring a minimal amount of new physics---would be the hardest to detect via direct observation. There are, however, some experimental protocols suggesting that this could be possible in the near future \cite{Schumann:2019eaa, Carney:2019pza, Perez:2023tld}.

Let us briefly review such {\em gravitational miracle} ignoring factors of order one (like factors of $\pi$, $\xi (3)$, etc) as they do not change the estimates and considerable simplify the presentation. 
Particles decouple from thermal equilibrium when their interaction rate $\Gamma$ drops below the Hubble rate $H$, namely when \cite{Peter:1208401}
\be\label{rating}
\Gamma\equiv n \sigma v< H,
\ee
where $n$ is the particle number density including all species, $\sigma$ its cross section of interaction, and $v$ the speed of the particles of interest with respect to the rest frame of the cosmological fluid that 
is assumed to be in thermal equilibrium. The Hubble rate can be estimated from the Friedmann equation using that the energy density $\rho\propto g_s T^4$, with $g_s$ the number of species involved, namely 
\be
H\approx 
\sqrt{g_s}  \frac{T^2}{m_p} .
\ee
The density $n\approx g_s T^3$, $\sigma\approx m_{pbh}^{2}/m_p^4$ (the PBH area) and we initially assume $v\approx 1$. Thus, we get
\be
\Gamma\approx \frac{g_sT^3 m_{pbh}^2}{m_p^4}
\ee
from which it follows that the decoupling temperature---below which \eqref{rating} is satisfied---is $
T_D\approx \frac {m_p} {\sqrt{g_s}} \left({m_{p}}/{m_{pbh}}\right)^2$. Thus, this first estimate of $T_D$ turns out to be compatible with non relativistic PBHs (contradicting $v\approx 1$). We can improve the estimate using that $v\approx (2T/m_{pbh})^{1/2}$ 
for non relativistic particles from which is follows that 
\be T_D\approx \frac{1}{(2g_s)^{1/3}}\frac{m_p}{m_{pbh}} m_p.\ee

The abundance fraction of Planckian BHs at decoupling relative to the total density is given by the corresponding Gibbs factor, namely 
\be
F\equiv \frac{\rho_{DM}(T_D)}{\rho(T_D)}\approx \exp\left(-\frac{m_{pbh}}{T_D}\right)\approx \exp\left(-{{(2g_s)}^{\frac13}} \frac{  m_{pbh}^2}{  m_p^2}\right). 
\ee  To predict how much dark matter of this type would remain around today we use that for $T\le T_D$ 
these BHs  behave like a pressure-less fluid decaying as $1/a^3$ or equivalently as $T^3$. This implies that 
the density of DM  goes like
\be
\rho_{DM}(T)={\rho_{DM} (T_D)} \frac{T^3}{T_D^3}\approx   \rho(T_D)  F \frac{T^3}{T_D^3}  \approx \underbrace{{{g_s}^{\frac23}}  \frac{T^3}{m_p^3} \frac{m_p} {m_{pbh}} \exp\left(-{(2g_s)^{\frac13}} \frac{  m_{pbh}^2}{  m_p^2}\right) }_{{\rm Need \ this \ factor \ order \ 10^{-120}\  at}\  T=T_{\rm today} } m_p^4.
\ee
Dark matter density today is about $10^{-120} m_p^4$. Using that $T^3_{\rm today} /T^3\approx 10^{-95}$ and assuming for concreteness $g_s\approx 100$ (this would be correct if the SM holds all the way to the Planck scale) one gets  $m_{pbh}\approx m_p$ which realizes the miracle we evoke. The dependence on $g_s$ is weak given the uncertainties in the matter sector beyond the standard model \footnote{\label{foot} As $g_s$ ranges from $427/4$ (the SM value) to, say, $2000$, the mass $m_{pbh}$ ranges from $3.15\  m_p$ to $1.97 \ m_p$! While $T_D$ ranges from $5\times 10^{-2} m_p\approx 50 \ T_{\rm \va GUT}$ to $3 \times 10^{-2} m_p\approx 30 \ T_{\rm \va GUT}$, i.e. scales not extraordinarily higher than usual precluding possible conflict with the negative evidence of the presence topological defects in cosmology.}. 

\section{Conclusions}\label{conclusions}

In conclusion, discreteness at the Planck scale implies that at Planckian curvatures the symmetries of the macroscopic universe we observe cannot be maintained. If symmetries are fundamentally broken and if a process of exponential expansion is at play, then the stochastic fluctuations of the fundamental quantum gravity scale induces inhomogeneities in the matter distribution that evolve into a scale invariant spectrum of perturbations as observed in the CMB. In such a paradigm the origin of structure is sourced in the preexistent structure of quantum geometry and not in vacuum fluctuations of trans-Planckian modes. The analysis presented here assumes exact de Sitter inflation. A proper account including slow-roll parameters will be available in \cite{gabrieles2}.

Due to the large number of microscopic degrees of freedom involved in the generation of the scalar inhomogeneities, a gaussian primordial semiclassical spectrum of perturbations is expected and standard arguments of the generation of non-gaussianities do not apply. As degrees of freedom violating scale invariance are primarily sensitive to the fundamental scale, gravitational waves are produced at a next to leading order in perturbation theory and are hence severely suppressed.
This invalidates the standard argument against a Planck scale inflationary scenario based on the negative observation of B modes in the CMB \cite{BICEP2:2018kqh}.   
The initially Planckian energy density of the universe---the most natural scale from the quantum gravity perspective---decays during a reheating phase that produces a hot big bang of a temperature near the Planck temperature. This leads to the thermal production of Planckian primordial BHs which, if stable, would produce the correct order of magnitude of dark matter. The view we propose in this letter is certainly in great contrast with the standard one. One should however keep in mind that, in spite of its long history, there is a large degree of speculation in the standard account of inflation as well. The value of the perspective opened here is, in our view, the possibility of looking at the observational facts from a different angle. This could show useful in making progress in understanding the mysterious features of present cosmology---ranging from the origin of structure, to the fundamental nature of dark matter, and dark energy---uncovering an unforeseen window linking the large scale structure of our universe with properties of the fundamental theory unifying gravity and quantum mechanics.

\begin{acknowledgments} 
We thank N. Bernal, and D. Sudarsky for discussions.
G.R.B is funded by project PIP 112-2021-0100225-CO of CONICET (Argentina). G.L. is funded by projects Universidad Nacional de La Plata I+D G175, and PIP 112-2020-0100729-CO of CONICET, and CONICET (Argentina). A.P. is funded by ID\#62312 grant from the John Templeton Foundation via the QISS Project (\href{qiss.fr}{qiss.fr}).
\end{acknowledgments}



\end{document}